\documentclass[12pt,amssymb,epsf]{article}
\usepackage{epsfig}
\newcommand{\gev}{\mbox{GeV}}
\newcommand{\mev}{\mbox{MeV}}
\newcommand{\real}{{\cal\mbox{Re\,}}}

\newcommand{\be}{\begin{equation}}
\newcommand{\ee}{\end{equation}}

\newcommand{\bbeta}{\bar{\beta}}
\newcommand{\VS}{\mbox{\tiny VS}}
\newcommand{\brem}{\mbox{\scriptsize brem}}
\newcommand{\res}{\mbox{\scriptsize res}}
\newcommand{\peak}{\mbox{\scriptsize peak}}

\begin{document}
\pagestyle{empty}

\begin{flushright}
{}
\end{flushright}
\vspace*{5mm}
\begin{center}
    {\bf FINAL-STATE RADIATION AND LINE-SHAPE DISTORTION \\[3mm]
       IN RESONANCE PAIR PRODUCTION} \\
\vspace*{1cm}
{\bf W.~Beenakker}$^{*)}$,\ \ 
{\bf F.A.~Berends} \ \  
{\bf and}  \ \ 
{\bf A.P.~Chapovsky}$^{\dagger)}$\\
\vspace{0.3cm}
Instituut--Lorentz, University of Leiden, The Netherlands
\vspace{2cm}\\
                                {\bf ABSTRACT} \\
\end{center}
\vspace*{5mm}
\noindent
In this letter it is shown how final-state QED corrections to the production
of a pair of resonances can distort the line shape of such a resonance in a
sizeable way. This effect depends on the definition of the line 
shape and can reach up to 30\%, depending on the final state. 
The mechanism is first displayed for a particular case of $ZZ$ production, for
which an exact and approximate treatment can be given. The approximate method 
is then applied to $W$-pair production. In addition some simple rules of thumb 
are given for accurately estimating the characteristic distortion effects,
like the mass shift and peak reduction.
\vspace*{5mm}
\vspace*{5cm}\\
\begin{flushleft}
May 1998
\end{flushleft}
\noindent
\rule[.1in]{16.5cm}{.002in}

\noindent
$^{*)}$Research supported by a fellowship of the Royal Dutch Academy of
Arts
and Sciences.\\
$^{\dagger)}$Research supported by the Stichting FOM.
\vspace*{0.3cm}

\vfill\eject

\setcounter{page}{1}
\pagestyle{plain}

\newpage

\section{Introduction}

As is well known \cite{zreport}, the $Z$ line shape as measured in
$e^{+}e^{-}\to Z\to 2f$ is distorted due to initial-state radiation (ISR).
Without ISR the total cross-section $\sigma(s)$ as a function of the
square of the centre-of-mass energy $s$ gives the line shape. With ISR the
centre-of-mass energy available to produce a $Z$ boson changes and, as a
consequence, so does the
shape of the total cross-section $\sigma(s)$. Experimentally the latter is
measured. If one would measure the square of the modified centre-of-mass 
energy $s'$, one would determine $\sigma(s')$ and thereby the pure line-shape.
It should be noted that final-state radiation (FSR) only marginally corrects 
the overall size of $\sigma(s)$, but not its shape. Therefore FSR is less
relevant for the usual $Z$ line-shape measurement.

When one produces two resonances, or one resonance and a stable particle, the 
line shape of such a resonance will be measured from the invariant-mass 
distribution of its decay products. Examples are pair production
of $W$ bosons, $Z$ bosons, $t\bar{t}$ or $HZ$. Depending on how one
measures the invariant-mass distribution of the decay products of the
particular resonance, one finds the pure line shape or a distorted one.
This time also FSR can cause the distortion.

It is the main purpose of this letter to point out that such a FSR-induced
distortion can arise. For exhibiting the effect we take an example for which 
we can perform both exact and approximate calculations. An ideal example is the
double-resonance process
\be
\label{eq:process}
 \nu_{\mu}\bar{\nu}_{\mu} \to ZZ
 \to e^{+}e^{-}\nu_{\tau}\bar{\nu}_{\tau}.
\ee
Here QED corrections apply only to the decay $Z\to e^{+}e^{-}$ and not to the 
other $Z$ decay or the initial state. When the $Z$ line shape is obtained from 
measuring the invariant-mass distribution of the $e^{+}e^{-}$ pair, FSR will 
distort it in a way reminiscent of the usual ISR distortion in single $Z$ 
production. The virtue of the example is threefold. In the first place, 
process~(\ref{eq:process}) is free of the gauge-invariance problems that are 
inherent 
in the production of unstable particles. This holds in spite of the fact that
we have left out all non-double-resonant mechanisms for producing the 
$e^{+}e^{-}\nu_{\tau}\bar{\nu}_{\tau}$ final state. Secondly, the QED
radiative corrections only lead to FSR. So, the effect of FSR on the line shape
can be studied without the additional presence of ISR phenomena. Thirdly, the 
effect can be calculated exactly. In more realistic examples, involving for 
instance $e^{+}e^{-}$ initial states, $Z$-pair production with both $Z$ bosons 
decaying into charged particles, or $W$-pair production, additional classes of 
QED radiative corrections emerge, like ISR \cite{isr} or non-factorizable 
interference
corrections \cite{nf-corr}. Moreover, in order to avoid gauge-invariance 
problems, the QED corrections often have to be calculated in an approximation,
which for instance restricts the calculation to the leading logarithmic 
corrections and/or the leading terms in a pole-scheme expansion around the 
resonances \cite{isr,pole-scheme}. 
Nevertheless the FSR distortion of the line shape will 
still be one of the main features. In these more complicated cases Monte Carlo
studies including radiative corrections would be needed. Here we focus 
exclusively on the line-shape deformation and its impact on the determination 
of the resonance mass.

Although we start with reaction~(\ref{eq:process}), we shall also
comment on the more realistic case of the $W$ line shape at LEP2.

\section{\boldmath The $Z$-pair example: exact calculation}

For process~(\ref{eq:process}) we first consider the Born approximation, to 
which two double-resonant diagrams contribute. After integration over the $Z$ 
production angle and the fermion decay angles, one obtains
\be
\label{eq:zz/born}
 \frac{d\sigma_{0}(M_1^2,M_2^2)}{dM_{1}^{2}\,dM_{2}^{2}}
 =
 \Pi(M_1^2,M_2^2)\,
 \frac{\Delta_{1}(M_{1}^2)}{|D_{1}(M_{1}^2)|^{2}}\,
 \frac{\Delta_{2}(M_{2}^2)}{|D_{2}(M_{2}^2)|^{2}}
 =
 F(M_{1}^{2}, M_{2}^{2}),
\ee
where $M_1^2$ and $M_2^2$ denote the invariant masses of the $e^+e^-$ and 
$\nu_{\tau}\bar{\nu}_{\tau}$ pairs, respectively. The spin-averaged production 
cross-section takes the form
\be
 \Pi(M_{1}^2,M_{2}^2)
 =
 \frac{G_F^2 M_Z^4}{4\pi s}\,(2g_{\nu})^4\,
 \frac{\sqrt{\lambda}}{s}\,
 \Biggl[
 -2 +
 \frac{s^2+(M_1^2+M_2^2)^2}{(s-M_1^2-M_2^2)\sqrt{\lambda}}
 \ln\biggl(\frac{s-M_1^2-M_2^2+\sqrt{\lambda}}{s-M_1^2-M_2^2-\sqrt{\lambda}}
    \biggr)
 \Biggr],
\ee
with $\lambda$ the Kallen function
\be
 \lambda = s^2+M_1^4+M_2^4-2\bigl(s M_1^2+sM_2^2+M_1^2 M_2^2\bigr),
\ee
which is in agreement with the literature \cite{zzborn}. The decay parts are 
given by
\begin{eqnarray}
 \Delta_{1}(M_{1}^2) &=&
     \frac{G_F M_Z^2}{6\pi\sqrt{2}}\,(g_{V\ell}^2+g_{A\ell}^2)\,\frac{1}{\pi}\,
     M_{1}^{2} \ =\ \frac{1}{\pi}\,M_{1}^{2}\,\frac{\Gamma_{Z\to e^+e^-}}{M_Z},
     \nonumber \\[1mm]
 \Delta_{2}(M_{2}^2) &=&
     \frac{G_F M_Z^2}{6\pi\sqrt{2}}\,2 g_{\nu}^2\,\frac{1}{\pi}\,M_{2}^{2} 
     \ =\ \frac{1}{\pi}\,M_{2}^{2}\,
     \frac{\Gamma_{Z\to \nu_{\tau}\bar{\nu}_{\tau}}}{M_Z},
\end{eqnarray}
whereas the resonance shapes are dominated by
\be
 D_{1,2}(M_{1,2}^{2}) = M_{1,2}^{2} - M_{Z}^{2} 
                        + i M_{1,2}^{2}\,\frac{\Gamma_{Z}}{M_Z}.
\ee
Note that we have used the standard LEP1 representation in the above formulae, 
involving $G_F$ and the effective couplings of the $Z$ boson to leptons
($g_{V\ell},\,g_{A\ell}$) and neutrinos ($g_{\nu}$).

Applying virtual and soft photonic corrections to~(\ref{eq:zz/born}) yields
\be
\label{eq:zz/virt}
 \frac{d\sigma_{\VS}(M_{1}^2,M_{2}^2)}{d M_{1}^{2}\,dM_{2}^{2}}
 =
 \frac{d\sigma_{0}(M_{1}^2,M_{2}^2)}{d M_{1}^{2}\,dM_{2}^{2}}
 \Biggl[
 1+\frac{2\alpha}{\pi}\,(L-1)\ln\epsilon
 +\frac{\alpha}{\pi}\biggl(\frac{3}{2}L+\frac{\pi^{2}}{3}-2\biggr)
 \Biggr],
\ee
with
\be
 L=\ln\biggl(\frac{M_{1}^{2}}{m_{e}^{2}}\biggr).
\ee
Here we have defined the soft photons in the rest frame of the $Z$: 
$E_{\gamma}<\epsilon M_{1}/2\ll\Gamma_{Z}$. Photon bremsstrahlung involving
more energetic photons introduces an explicit dependence on the photon energy 
$E_{\gamma}$, resulting in a distribution in the invariant masses of both the
$e^+e^-$ pair ($\tilde{M}_1^2$) and the $e^+e^-\gamma$ system 
($M_1^2 =$ virtuality of the $Z$ boson):
\be
\label{eq:zz/hard}
 \frac{d\sigma_{\brem}(M_{1}^2,\tilde{M}_{1}^{2},M_{2}^2)}
      {d M_{1}^{2}\,d \tilde{M}_{1}^{2}\,d M_{2}^{2}}
 =
 \frac{d\sigma_{0}(M_{1}^2,M_{2}^2)}{dM_{1}^{2}\,d M_{2}^{2}}\,
 \frac{\alpha}{\pi}\,
 \bigl(L'-1\bigr)\,
 \frac{1+z^{2}}{1-z}\,\frac{1}{M_{1}^{2}},
\ee
where
\be
 z=\frac{\tilde{M}_{1}^{2}}{M_{1}^{2}}=\frac{1}{\zeta},
 \ \ \
 L'=\ln\biggl(\frac{\tilde{M_{1}^{2}}}{m_{e}^{2}}\biggr)
 =
 L+\ln z.
\ee
When correction~(\ref{eq:zz/virt}) is combined with~(\ref{eq:zz/hard})
and an integration over $\tilde{M}_{1}^{2}$ is performed, the correction
to $d\sigma_{0}/(d M_{1}^{2}\,d M_{2}^{2})$ takes on the form of the usual FSR
factor $1+3\alpha/(4\pi)$. This is in agreement with the KLN theorem, which
implies that the large logarithmic contributions ($\propto L,\,L'$) vanish 
upon summation (integration) over all degenerate final states.
So, the resonance shape is not deformed when one measures the 
$M_{1}^{2}$ distribution, i.e.~the invariant-mass distribution of the
$e^{+}e^{-}\gamma$ system.

In our special example this choice of distribution is, of course, the natural
one. However, in more realistic processes it is in general unclear whether
the photon is radiated from the initial state, the unstable particles, or the
final state. This introduces the freedom to either choose $M_1^2$ or 
$\tilde{M}_1^2$ for the definition of the invariant-mass distribution of the
unstable particle. 

If one measures the $\tilde{M}_{1}^{2}$ distribution
one will find a distorted line shape. The reason is that~(\ref{eq:zz/hard})
now has to be integrated over $M_{1}^2$ values ranging from $\tilde{M}_{1}^2\,$
to $\,(\sqrt{s}-M_{2})^2$. This causes the $\tilde{M}_{1}^{2}$ line shape to
receive contributions from effectively {\it higher} $Z$-boson virtualities. 
This is to be compared with the single-$Z$-production case where the 
ISR-corrected line shape receives contributions from effectively {\it lower} 
$Z$-boson virtualities. Due to the fact that roughly speaking the resonance
shape is symmetric around the resonance mass, one expects now a distortion of 
the resonance shape that is approximately the LEP1 distortion reflected with
respect to the resonance mass.

The bremsstrahlung contribution to the line shape
$d\sigma/(d\tilde{M}_{1}^{2}\,d M_{2}^{2})$ arising from~(\ref{eq:zz/hard}) 
reads
\be
\label{eq:zz/hard2}
 \frac{d\sigma_{\brem}(\tilde{M}_{1}^{2},M_{2}^2)}
      {d \tilde{M}_{1}^{2}\,d M_{2}^{2}}
 =
 \frac{\alpha}{\pi}\,(L'-1)
 \int\limits_{1+\epsilon}^{\zeta_{\max}}
 d \zeta\,
 F(\zeta \tilde{M}_{1}^{2}, M_{2}^{2})
 \Biggl[
 \frac{2}{\zeta-1}-\frac{1+\zeta}{\zeta^{2}}
 \Biggr],
\ee
where the function $F$ is defined in~(\ref{eq:zz/born}) and 
$\zeta_{\max}=(\sqrt{s}-M_{2})^{2}/\tilde{M}_{1}^{2}$. 
Combining~(\ref{eq:zz/virt}) and~(\ref{eq:zz/hard2}) 
gives the ${\cal O}(\alpha)$ correction to the line shape. 
Just like at LEP1 it will be necessary to resum the soft corrections, as will
become clear from the discussion in the following section. Based on LEP1 
experience \cite{zreport} a suitable expression for this resummation is given 
by 
\be
\label{eq:zz/res}
 \frac{d \sigma_{\res}(\tilde{M}_{1}^{2},M_{2}^2)}
      {d\tilde{M}_{1}^{2}\,d M_{2}^{2}}
 =
 \int\limits_{1}^{\zeta_{\max}} d\zeta\,G(\zeta)\,
 F(\zeta\tilde{M}_{1}^{2}, M_{2}^{2}),
\ee
with
\begin{eqnarray}
 G(\zeta)     &=& \beta\,(\zeta-1)^{\beta-1}
                  (1+\delta^{\VS}_{\res})
                  - \frac{\beta}{2}\,\frac{1+\zeta}{\zeta^{2}} \nonumber \\
 \beta        &=& \frac{2\alpha}{\pi}\,(L'-1) \nonumber \\
 \delta^{\VS}_{\res} &=& \frac{\alpha}{\pi}\biggl(
                  \frac{3}{2}L'+\frac{\pi^2}{3}-2\biggr).
\end{eqnarray}

\begin{figure}[t]
  \unitlength 1cm
  \begin{center}
  \begin{picture}(13.5,6.5)
  \put(0.5,4){\makebox[0pt][c]
            {\boldmath $\frac{d\sigma}{d\tilde{M}_{1}^{2}d M_{2}^{2}}$}} 
  \put(0.5,3){\makebox[0pt][c]
            {\boldmath $\bigl[\mbox{\bf\scriptsize pb}/
             \mbox{\scriptsize\bf GeV}^{4}\bigr]$}}
  \put(11.5,-0.3){\makebox[0pt][c]{\boldmath $\tilde{M}_{1}\,
                  [\mbox{\bf \scriptsize GeV}]$}}
  \put(1,-5){\includegraphics{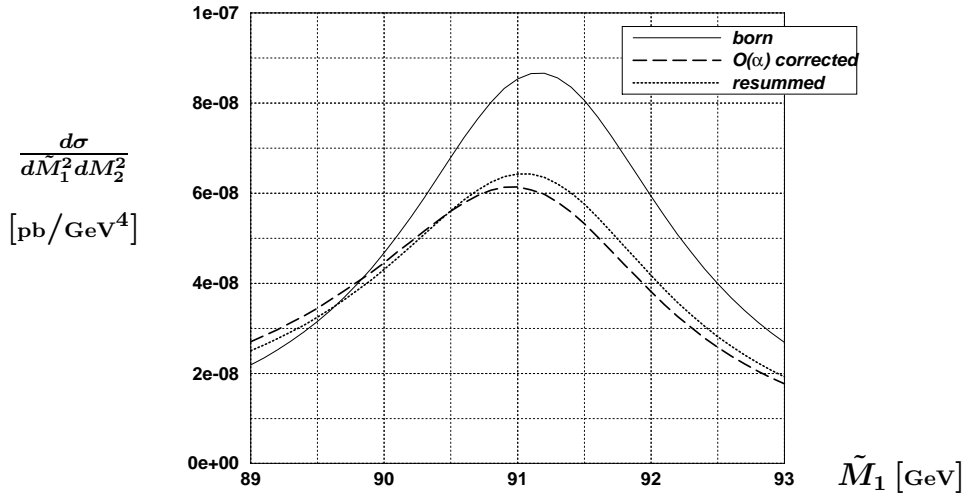}}
  \end{picture}
  \end{center}
  \caption[]{The FSR-induced distortion of the line shape
             $d\sigma/(d \tilde{M}_{1}^{2}\,d M_{2}^{2})$ corresponding to
             process~(\protect\ref{eq:process}) for $M_2=M_Z$. Centre-of-mass
             energy: $\protect\sqrt{s}=200\,\gev$.}
\label{fig:1}
\end{figure}

In Fig.~\ref{fig:1} we display the FSR-induced distortion effects on the line 
shape $d\sigma/(d\tilde{M}_{1}^{2}\,d M_{2}^{2})$ for a centre-of-mass energy
of $\sqrt{s}=200\,\gev$ and a fixed invariant mass $M_2=M_Z$. However, the 
actual distortion phenomena do not depend on the precise value of $M_2$. The
parameter input used in the numerical evaluation is: 
\begin{eqnarray*}
  M_Z\!\!\!&=&\!\!\! 91.1867\,\gev,\ \Gamma_Z=2.4948\,\gev, 
  \ G_F=1.16639\times 10^{-5}\,\gev^{-2},\ m_e=0.51099906\,\mev, \\
  \alpha^{-1}\!\!\!&=&\!\!\! 137.0359895,\ g_{\nu}=0.50125,\  
  g_{V\ell}=-0.03681,\ g_{A\ell}=-0.50112.
\end{eqnarray*}
 The sizeable distortion 
effects are clearly visible, just as the importance of the soft-photon 
resummation. Compared with the Born line shape, the ${\cal O}(\alpha)$ 
(resummed) QED corrections induce a shift in the peak position of $-199\,\mev$
($-112\,\mev$) and a reduction of the peak height by 29\% (26\%). The size of
these effects are a direct result of the non-cancellation of the leading 
logarithmic corrections, which can be understood from the observation that a 
fixed value for $\tilde{M}_1$ makes it impossible to sum over all 
degenerate final states. Another noteworthy observation is the close 
similarity of the curves in Fig.~\ref{fig:1} to the ones for the Z line shape 
at LEP1 \cite{zreport}. As predicted, the two sets of curves are approximately 
related by reflection with respect to the Born peak position.

\section{\boldmath The $Z$-pair example: approximations}

As mentioned before, in order to calculate QED corrections to more
realistic processes like $W$-pair production one in general has to resort to
approximations. First of all, the fact that we are dealing with unstable 
(charged) particles introduces the problem of a gauge-invariant treatment of 
the finite-width effects \cite{isr}. An appropriate way of handling this 
problem is by applying the pole scheme, i.e.~by performing an expansion around 
the resonances. When it comes to ${\cal O}(\alpha)$ corrections, it is 
sufficient to consider only the leading (double-pole) term in this expansion,
leaving out terms that are formally suppressed by at least
$\alpha L\,\Gamma/(\pi M)$. From now on we will refer to this procedure as the
double-pole approximation and indicate quantities that are calculated in this 
approximation by a bar. Note that the approximation only makes sense near the 
resonance of the unstable particle and sufficiently far above the production 
threshold of the underlying on-shell production process. The latter is caused 
by the direct relation between the double-pole residues and the on-shell 
production and decay processes. Secondly, as we have seen in the previous 
section, the leading logarithmic corrections ($\propto L'$) 
constitute the bulk of the FSR distortion effects. Moreover, these leading-log
effects are universal and gauge invariant, being directly related to the 
collinear limit of photon radiation off light particles (like~$e^{\pm}$). In
particular the universality property is appealing, since it implies that the 
description of the leading-log corrections does not depend on the specific 
features of the unstable particles and their photonic interactions. Therefore, 
it is worthwhile to further restrict the double-pole calculations to the 
leading logarithms. This additional approximation is referred to as the 
leading-log approximation (LLA). Before making any comments on processes like 
$W$-pair production, we first concentrate on our $Z$-pair example and check 
the validity of the indicated approximations. 
 
We start off with the definition of the double-pole approximation. At Born 
level it amounts to 
\be
 \frac{d\bar{\sigma}_{0}(M_1^2,M_2^2)}{dM_{1}^{2}\,dM_{2}^{2}}
 =
 \Pi(M_Z^2,M_Z^2)\,
 \frac{\Delta_{1}(M_{Z}^2)}{|\bar{D}_{1}(M_1^2)|^{2}}\,
 \frac{\Delta_{2}(M_{Z}^2)}{|\bar{D}_{2}(M_2^2)|^{2}},
\ee
with 
\be
 \bar{D}_{1,2}(M_{1,2}^{2}) = M_{1,2}^{2}-M_{Z}^{2} + i M_Z\Gamma_{Z}.
\ee

As mentioned before, the QED radiative corrections introduce an ambiguity in
defining the invariant-mass distributions of the unstable particles. The most
transparent way of illustrating this is by considering the case that the
photon is radiated from such an unstable particle. For the description of the 
resonance before (after) radiation the natural choice of invariant mass 
involves the fermion pair with (without) the photon. The pole expansion can 
now in principle be performed around either resonance. In practice one has to 
choose one particular invariant mass for the distributions. For the purpose of
studying FSR-induced distortion effects, we shall choose the
$e^+e^-$ invariant mass $\tilde{M}_{1}$ in the following, although the
$e^+e^-\gamma$ invariant mass $M_1$ would have been more natural for our 
special example~(\ref{eq:process}).
The corresponding double-pole approximation forces us to replace 
$M_1^2=\zeta \tilde{M}_1^2$ by $\zeta M_Z^2$, introducing an explicit 
dependence on the photon energy in the double-pole residues. This would even 
affect the (neutral) resonance-pair-production stage of the process. 
However, as can be verified explicitly, only semi-soft photons with
energy $E_{\gamma}={\cal O}(\Gamma_Z)\ll M_Z$ contribute to the 
${\cal O}(\alpha)$ corrected double-pole residues. As a result, $\zeta$ can be 
effectively replaced by unity whenever possible, re-establishing the usual 
form of the double-pole approximation in terms of off-shell Breit--Wigner
distributions and on-shell production/decay processes. The effects from hard 
photons ($E_{\gamma}\gg\Gamma_Z$) are suppressed by at least $\Gamma_Z/M_Z$ and
are therefore neglected in the double-pole approximation. 
The picture underlying this phenomenon is that hard photons move the $Z$-boson 
virtuality ($M_1^2$) far off resonance for near-resonance $\tilde{M}_1^2$ 
values, resulting in a suppressed contribution to the $\tilde{M}_1^2$ line 
shape. In fact, only the (soft) $1/(\zeta-1)$ term in~(\ref{eq:zz/hard2}) 
contributes to the $\tilde{M}_1^2$ line shape in the double-pole approximation.
It should be noted that this very suppression of hard-photon effects serves
as {\it a posteriori} justification of the soft-photon resummation proposed 
in~(\ref{eq:zz/res}). 

With this observation in mind, the bremsstrahlung 
contribution~(\ref{eq:zz/hard2}) takes the following form in double-pole 
approximation:
\be 
 \frac{d \bar{\sigma}_{\brem}(\tilde{M}_{1}^{2},M_{2}^2)}
      {d\tilde{M}_{1}^{2}\,d M_{2}^{2}}
 =
 \frac{d\bar{\sigma}_{0}(\tilde{M}_1^2,M_2^2)}{d\tilde{M}_{1}^{2}\,dM_{2}^{2}} 
 \int\limits_{1+\epsilon}^{\infty} d \zeta\,\frac{\bbeta}{\zeta-1}\,
 \frac{|\bar{D}_1(\tilde{M}_1^2)|^2}{|\bar{D}_1(\zeta\tilde{M}_1^2)|^2}.
\ee
Here $\bbeta$ can be derived from $\beta$ by setting $\tilde{M}_{1}^{2}=M_Z^2$.
Note that the upper integration boundary 
$\zeta_{\max}$ has been extended to infinity, which is motivated by the fact 
that hard-photon effects are sufficiently suppressed. The remaining integral
can be performed analytically. Combining with the virtual and soft corrections,
which can be readily derived from~(\ref{eq:zz/virt}), we obtain in the 
double-pole LLA 
\be
 \frac{d \bar{\sigma}(\tilde{M}_{1}^{2},M_{2}^2)}
      {d\tilde{M}_{1}^{2}\,d M_{2}^{2}}
 =
 \frac{d\bar{\sigma}_{0}(\tilde{M}_1^2,M_2^2)}{d\tilde{M}_{1}^{2}\,dM_{2}^{2}} 
 \Biggl\{ 
 1 + \frac{3}{4}\,\bbeta 
   + \bbeta\,\real\Biggl[\frac{i\bar{D}_1^*(\tilde{M}_1^2)}{M_Z\Gamma_Z}
     \ln\biggl(\frac{\bar{D}_1(\tilde{M}_1^2)}{M_Z^2}\biggr)\Biggr]
 \Biggr\}.
\ee
In a way similar to the previous section the soft-photon corrections can be
resummed, but this time the integral can be carried out explicitly in the
double-pole LLA:
\begin{eqnarray}
\label{LLA:resummed}
 \frac{d \bar{\sigma}_{\res}(\tilde{M}_{1}^{2},M_{2}^2)}
      {d\tilde{M}_{1}^{2}\,d M_{2}^{2}}
 &=&
 \frac{d\bar{\sigma}_{0}(\tilde{M}_1^2,M_2^2)}{d\tilde{M}_{1}^{2}\,dM_{2}^{2}}
 \,\bigl(1+\frac{3}{4}\,\bbeta)
 \int\limits_{1}^{\infty} d\zeta\,\bbeta\,(\zeta-1)^{\bbeta-1}\,
 \frac{|\bar{D}_1(\tilde{M}_1^2)|^2}{|\bar{D}_1(\zeta\tilde{M}_1^2)|^2}
 \nonumber \\[1mm]
 &=&
 \frac{d\bar{\sigma}_{0}(\tilde{M}_1^2,M_2^2)}{d\tilde{M}_{1}^{2}\,dM_{2}^{2}}
 \,\bigl(1+\frac{3}{4}\,\bbeta)
 \,\frac{\pi\bbeta}{\sin(\pi\bbeta)}\,
 \real\Biggl[\frac{i\bar{D}_1^*(\tilde{M}_1^2)}{M_Z\Gamma_Z}\,
      \biggl(\frac{\bar{D}_1(\tilde{M}_1^2)}{M_Z^2}\biggr)^{\bbeta}\,\Biggr]. 
\end{eqnarray}

Having calculated the same quantities as in the previous section, we are now in
the position to check the validity of the double-pole LLA. It turns out that
the approximated results exhibit the same FSR distortions as 
the exact ones. Upon closer investigation, we observe for the 
${\cal O}(\alpha)$ (resummed) QED corrections a shift in the peak position of 
$-193\,\mev$ ($-113\,\mev$) and a reduction of the peak height by 29\% (26\%).
This is in excellent agreement with the distortion parameters of the exact
calculation, proving the viability of the adopted approximations.\footnote{
  For completeness we note that all curves in the double-pole LLA are 
  displaced by a small amount with respect to the exact ones. This is caused 
  by the fact that the
Born results differ by subleading terms in the pole expansion.}%

Based on the results in the double-pole LLA, it is possible to derive simple
and sufficiently accurate rules of thumb for the distortion parameters:
\begin{itemize}
\item ${\cal O}(\alpha)$ corrections: the shift in the peak position
      $\Delta\,\tilde{M}_1^{\peak}$ and the corresponding peak
      reduction factor $\kappa^{\peak}$ with respect to the Born
      line shape can be approximated by 
\begin{eqnarray} 
  \Delta\,\tilde{M}_1^{\peak} &\approx& 
      - \,\frac{\pi\bbeta\,\Gamma_Z/8}{\kappa^{\peak}-3\bbeta/2}
      \ =\  -196\,\mev, \nonumber \\
  \kappa^{\peak} &\approx& 
      1 + \bbeta\,\ln\biggl(\frac{\Gamma_Z}{M_Z}\biggr) + \frac{3}{4}\,\bbeta
    + \frac{\pi^2}{16}\,\bbeta^2 \ =\ 0.70;
\end{eqnarray}
\item resummed corrections: now the distortion parameters read
\begin{eqnarray}
\label{thumb:res}
  \Delta\,\tilde{M}_1^ {\peak} &\approx&
      - \frac{\pi}{8}\,\bbeta\,\Gamma_Z (1 + \frac{\bbeta}{2}) \ =\ -111\,\mev,
      \nonumber \\
  \kappa^{\peak} &\approx& 
  \biggl(\frac{\Gamma_Z}{M_Z}\biggr)^{\bbeta} (1 + \frac{3}{4}\,\bbeta)\,
  (1 + \frac{5\pi^2}{48}\,\bbeta^2 - \frac{\pi^2}{32}\,\bbeta^3) \ =\ 0.74.
\end{eqnarray}
\end{itemize}
This is in perfect agreement with the observed exact and double-pole
distortion parameters. The analogy with the rules of thumb derived for the 
$Z$ line shape at LEP1 \cite{thumb} confirms the relation between the 
FSR-induced distortion effects in double $Z$-resonance production and the 
ISR-induced distortion effects in single $Z$-resonance production at LEP1.

\section{\boldmath Some comments on the $W$ line shape at LEP2}

As has been shown in the previous section, the double-pole LLA constitutes a 
reliable framework for a gauge-invariant and universal description of 
FSR-induced distortion phenomena in double-resonance production. The essence
of these phenomena is fully contained in the correction factor presented 
in~(\ref{LLA:resummed}), 
which applies to each individual distorted Breit--Wigner 
distribution. For two distorted distributions the effect is hence 
multiplicative. Consequently, the reduction factor for a double-invariant-mass
distribution is given by the product of the reduction factors for the
individual single-invariant-mass distributions. However, the shift in the
peak position does not change in the presence of more than one resonance; it
only depends on the decay products of the unstable particle that is 
investigated. The only differences between process~(\ref{eq:process}) and the 
more realistic process of $W$-pair production at LEP2 are the resonance
parameters ($M_W=80.22\,\gev$ and $\Gamma_W=2.08\,\gev$) and the fact that 
$\bbeta$ depends on the decay products of the decaying particle. For instance, 
the leptonic $W$ decays involve only {\it one} charged lepton instead of two. 
As a result, we should use~(\ref{LLA:resummed}) with 
$\bbeta \to \frac{\alpha}{\pi}\,[ \ln(M_W^2/m_{\ell}^2)-1 ]\,$ for 
$\,\ell = e,\mu,\tau$, which is scaled down by at least a factor of
two compared with the $Z$-pair example. For $W$~bosons decaying into an 
electron or positron, the resummed FSR distortion effects amount to a shift in
the peak position of $-45\,\mev$ and a peak reduction factor of 0.86 per
distorted resonance (i.e.~0.74 for a double-invariant-mass distribution), 
as can also be read off from~(\ref{thumb:res}).\footnote{
 In more realistic event-selection procedures also a
 minimum opening angle ($\theta_0$) between the lepton and photon might be
 required for a proper identification of both particles. The effect of this
 can be represented by using $\ln(4/\theta_0^2)$ instead of 
 $[\ln(M_W^2/m_{\ell}^2)-1]\,$ in the definition of $\bbeta$.}%

{}From the previous discussions it should be clear that FSR-induced distortion
effects can be sizeable and should be taken into account properly in the
Monte Carlo programs that are used for the $W$-mass determination at LEP2.

\end{document}